%% file: paper.tex
\documentclass{aa}
\usepackage{graphicx}
\usepackage{txfonts}
\usepackage{url}
\graphicspath{{./fig/}{./png/}}
\input aa

\begin{document}

\title{Scale dependence of alpha effect and turbulent diffusivity}
\author{A. Brandenburg\inst{1} \and K.-H. R\"adler\inst{2}
\and M. Schrinner\inst{3}}

\institute{
NORDITA, Roslagstullsbacken 23, SE-10691 Stockholm, Sweden
\and
Astrophysical Institute Potsdam, An der Sternwarte 16, D-14482 Potsdam, Germany
\and
Max-Planck-Institut f\"ur Sonnensystemforschung, 37191 Katlenburg-Lindau,
Germany
}

%\date{Received September 15, 1996; accepted March 16, 1997}
\date{\today}

\abstract{}{
To determine alpha effect and turbulent magnetic diffusivity for
mean magnetic fields with profiles of different length scale from
simulations of isotropic turbulence, and to relate these results
to nonlocal formulations in which alpha and the turbulent magnetic diffusivity
correspond to integral kernels.
}{
A set of evolution equations
for magnetic fields is solved which gives the response to imposed test fields,
that is, mean magnetic fields with various wavenumbers.
Both an imposed fully helical steady flow consisting of a pattern of screw-like motions (Roberts flow)
and time-dependent statistically steady isotropic turbulence are considered.
In the latter case the aforementioned evolution equations are solved simultaneously
with the momentum and continuity equations.
The corresponding results for the electromotive force are used to
calculate alpha and magnetic diffusivity tensors.
}{
For both the Roberts flow under the second--order correlation approximation
and isotropic turbulence
alpha and turbulent magnetic diffusivity are largest at large
scales and these values diminish toward smaller scales.
In both cases the alpha effect and turbulent diffusion kernels are well approximated
by exponentials, corresponding to Lorentzian profiles in Fourier space.
For isotropic turbulence the turbulent diffusion kernel is half as wide as the
alpha effect kernel.
For the Roberts flow beyond the second--order correlation approximation
the turbulent diffusion kernel becomes negative at large scales.
}{}
\titlerunning{Scale dependence of alpha effect}
\maketitle

\section{Introduction}
\label{Intro}

Stars and galaxies harbour magnetic fields whose scales are large compared
with the scale of the underlying turbulence.
This phenomenon is successfully explained in terms of mean--field dynamo theory
discussed in detail in a number of text books and reviews
(e.g.\ Moffatt 1978, Krause \& R\"adler 1980, Brandenburg \& Subramanian 2005a).
In this context velocity and magnetic fields are split into large--scale
and small--scale components, $\UU=\meanUU+\uu$ and $\BB=\meanBB+\bb$,
respectively.
The crucial quantity of the theory is the mean electromotive force caused by
the small--scale fields,
$\meanEMF=\overline{\uu\times\bb}$.
In many representations it is discussed under strongly simplifying assumptions.
Often the relationship between the mean electromotive force and the mean magnetic field
is tacitly taken as (almost) local and as instantaneous,
that is, $\meanEMF$ in a given point in space and time is considered as determined
by $\meanBB$ and its first spatial derivatives in this point only,
and the possibility of a small--scale dynamo is ignored.
Then the mean electromotive force is given by
\EQ
\meanemf_i=\alpha_{ij}\meanB_j+\eta_{ijk} \partial \meanB_j / \partial x_k
\label{meanemfi}
\EN
with two tensors $\alpha_{ij}$ and $\eta_{ijk}$.
If the turbulence is isotropic the two tensors are isotropic, too, that is
$\alpha_{ij}=\alpha \delta_{ij}$ and $\eta_{ijk}=\eta_{\rm t} \epsilon_{ijk}$
with two scalar coefficients $\alpha$ and $\eta_{\rm t}$.
Then the expression \eq{meanemfi} simplifies to
\EQ
\meanEMF=\alpha\meanBB-\eta_{\rm t}\meanJJ \, ,
\label{meanEMF}
\EN
where we have denoted $\nab \times \BB$ simply by $\JJ$
(so that $\JJ$ is $\mu_0$ times the electric current density,
where $\mu_0$ is the magnetic permeability of free space).
The coefficient $\alpha$ is, unlike $\eta_{\rm t}$, only
non--zero if the turbulence lacks mirror--symmetry.
The coefficient $\eta_{\rm t}$ is referred to as the turbulent
magnetic diffusivity.

In general, the mean electromotive force has the form
\EQ
\meanEMF=\meanEMF_0 + \KK \circ \meanBB,
\label{kernel}
\EN
where $\meanEMF_0$ stands for a part of $\meanEMF$ that is independent of $\meanBB$,
and $\KK \circ \meanBB$ denotes a convolution in space and time of a kernel $\KK$
with $\meanBB$ (see, e.g., Krause \& R\"adler 1980, R\"adler 2000, R\"adler \& Rheinhardt 2007).
Due to this convolution, $\meanEMF$ in a given point in space and time depends on $\meanBB$
in a certain neighborhood of this point, with the exception of future times.
This corresponds to a modification of \eq{meanemfi} such that also higher spatial
and also time derivatives of $\meanBB$ occur.

In this paper we ignore the possibility of coherent effects resulting from
small--scale dynamo action and put therefore $\meanEMF_0$
equal to zero.
For the sake of simplicity we assume further the connection between $\meanEMF$
and $\meanBB$ to be instantaneous so that the convolution $\KK \circ \meanBB$
refers to space coordinates only.
The memory effect, which we so ignore, has been studied
previously by solving an evolution equation for $\meanEMF$
(Blackman \& Field 2002).

For homogeneous isotropic turbulence we may then write,
analogously to (\ref{meanEMF}),
\EQ
\meanEMF=\hat{\alpha}\circ\meanBB-\hat{\eta}_{\rm t}\circ\meanJJ,
\label{meanEMFzz}
\EN
or, in more explicit form,
\EQ
\meanEMF(\xx)=\int\left[\hat{\alpha}(\xi)\meanBB(\xx-\xxi)
-\hat{\eta}_{\rm t}(\xi)\meanJJ(\xx-\xxi)\right]\,\dd^3 \xi
\label{meanEMFzzprime}
\EN
with two functions $\hat{\alpha}$ and $\hat{\eta}_{\rm t}$ of $\xi = |\xxi|$
that vanish for large $\xi$.
The integration is in general over all $\xi$--space.
Although $\meanEMF$ and $\meanBB$ as well as $\hat{\alpha}$ and
$\hat{\eta}_{\rm t}$ may depend on time
the argument $t$ is dropped everywhere.
For a detailed justification of the relations \eq{meanEMFzz} and \eq{meanEMFzzprime}
we refer to \App{justific}.
In the limit of a weak dependence of $\meanBB$ and $\meanJJ$ on space coordinates,
i.e.\ when the variations of $\meanBB (\xx - \xxi)$ and $\meanJJ (\xx - \xxi)$ with $\xxi$
are small in the range of $\xi$ where $\hat{\alpha} (\xi)$ and $\hat{\eta}_{\rm t} (\xi)$
are markedly different from zero, the relations \eq{meanEMFzz} or \eq{meanEMFzzprime}
turn into \eq{meanEMF},
and we see that \mbox{$\alpha = \int \hat{\alpha} (\xi) \, \dd^3 \xi$}
and $\eta_{\rm t} = \int \hat{\eta}_{\rm t} (\xi) \, \dd^3 \xi$.

At first glance the representations \eq{meanEMFzz} and \eq{meanEMFzzprime} of $\meanEMF$
look rather different from \eq{kernel}.
Considering $\meanJJ = \nab \times \meanBB$ and carrying out an integration by parts we may
however easily rewrite \eq{meanEMFzzprime} into
\EQ
\meanemf_i (\xx) = \int K_{ij} (\xxi) \meanB_j (\xx - \xxi) \, \dd^3\xi
\label{eq1}
\EN
with
\EQ
K_{ij} (\xxi) = \hat{\alpha} (\xi) \delta_{ij}
    - \frac{1}{\xi} \frac{\partial \hat{\eta}_{\rm t} (\xi)}{\partial \xi} \epsilon_{ijk} \xi_k \, .
\label{eq3}
\EN
We further note that due to the symmetry of $\hat{\alpha} (\xi)$ in $\xxi$
only the part of $\meanBB (\xx - \xxi)$ that is symmetric in $\xxi$,
i.e.\ the part that can be described by $\meanBB (\xx)$ and its derivatives of even order,
contributes to the $\hat{\alpha}$ terms in \eq{meanEMFzzprime} or in \eq{eq1} and \eq{eq3}.
The symmetry of $\hat{\eta}_{\rm t} (\xi)$ implies
that only the part of $\meanBB (\xx - \xxi)$ antisymmetric in $\xxi$,
which corresponds to the derivatives of $\meanBB (\xx)$ of odd order,
contributes to the $\hat{\eta}_{\rm t}$ terms.

Finally, referring to a Cartesian coordinate system $(x,y,z)$ we define mean fields
by averaging over all $x$ and $y$ so that in particular $\meanEMF$ and $\meanBB$
depend only on $z$ and on time.
Then \eq{meanEMFzzprime} turns into
\EQ
\meanEMF(z)
=\int\left[\hat{\alpha}(\zeta)\meanBB(z-\zeta)
-\hat{\eta}_{\rm t}(\zeta)\meanJJ(z-\zeta)\right]\,\dd \zeta \, .
\label{meanEMFzzprime2}
\EN
The functions $\hat{\alpha} (\zeta)$ and $\hat{\eta}_{\rm t} (\zeta)$
are just averages of $\hat{\alpha} (\xi)$ and $\hat{\eta}_{\rm t} (\xi)$ over all $\xi_x$ and $\xi_y$.
They are therefore real and symmetric in $\xi_z\equiv\zeta $.
The integration in \eq{meanEMFzzprime2} is in general over all $\zeta$.
The remark on the limit of weak dependences of $\meanBB$ and $\meanJJ$ on space coordinates
made above in connection with \eq{meanEMFzz} and \eq{meanEMFzzprime}
applies analogously to \eq{meanEMFzzprime2}.
We have now $\alpha = \int \hat{\alpha} (\zeta) \, \dd \zeta$
and $\eta_{\rm t} = \int \hat{\eta}_{\rm t} (\zeta) \, \dd \zeta$.

Relation \eq{meanEMFzzprime2} can also be brought in a form analogous to \eq{eq1} and \eq{eq3},
\EQ
\meanemf_i (z) = \int K_{ij} (\zeta) \meanB_j (z - \zeta) \, \dd \zeta
\label{eq5}
\EN
with
\EQ
K_{ij} (\zeta) = \hat{\alpha} (\zeta) \delta_{ij}
    - \frac{\partial \hat{\eta}_{\rm t} (\zeta)}{\partial \zeta} \epsilon_{ij3} \, .
\label{eq7}
\EN
The remarks made under \eq{eq1} and \eq{eq3} apply, now due to the symmetries
of $\hat{\alpha} (\zeta)$ and $\hat{\eta}_{\rm t} (\zeta)$ in $\zeta$,
analogously to \eq{meanEMFzzprime2}, \eq{eq5} and \eq{eq7}.

It is useful to consider in addition to \eq{meanEMFzzprime2} also the corresponding Fourier representation.
We define the Fourier transformation in this paper
by $Q(z) = \int \tilde{Q}(k) \exp(\ii k z) \, \dd (k/2 \pi)$.
Then this representation reads
\EQ
\tilde{\meanEMF}(k)
= \tilde{\alpha}(k) \tilde{\meanBB}(k)
-\tilde{\eta}_{\rm t}(k) \tilde{\meanJJ}(k) \, .
\label{meanEMFzzprime3}
\EN
Both $\tilde{\alpha}(k)$ and $\tilde{\eta}_{\rm t}(k)$ are real quantities,
and they are symmetric in $k$.
The limit of weak dependences of $\meanBB$ and $\meanJJ$ on $z$ corresponds here to $k \to 0$,
and we have $\alpha = \tilde{\alpha} (0)$ and $\eta_{\rm t} = \tilde{\eta}_{\rm t} (0)$.
Detailed analytic expressions for $\hat{\alpha}(\zeta)$ and $\hat{\eta}_{\rm t}(\zeta)$,
or $\tilde{\alpha}(k)$ and $\tilde{\eta}_{\rm t}(k)$, can be derived,
e.g., from results presented in Krause \& R\"adler (1980).
A numerical determination of quantities corresponding to
$\hat{\alpha} (\zeta)$ and $\hat{\eta}_{\rm t} (\zeta)$
has been attempted by Brandenburg \& Sokoloff (2002) for
shear flow turbulence.

In this paper two specifications of the velocity field $\uu$ will be considered.
In the first case $\uu$ is chosen such that it corresponds to a steady Roberts flow,
which is periodic in $x$ and $y$ and independent of $z$.
A mean--field theory of a magnetic field in fluid flows of this type, that are of course different
from genuine turbulence, has been developed in the context of the Karlsruhe dynamo experiment
(R\"adler et al.\ 2002a,b, R\"adler \& Brandenburg 2003).
It turned out that the mean electromotive force $\meanEMF$, except its $z$ component,
satisfies relation \eq{meanEMF}
if any nonlocality in the above sense is ignored (see also \App{robdyn}).
Several analytical and numerical results are available for comparison with those of this paper.
In the second case $\uu$ is understood as homogeneous, isotropic, statistically steady turbulence,
for which the above explanations apply immediately.
Employing the method developed by Schrinner et al.\ (2005, 2007) we will in both cases
numerically calculate the functions $\tilde{\alpha}(k)$ and $\tilde{\eta}_{\rm t}(k)$
as well as $\hat{\alpha}(\zeta)$ and $\hat{\eta}_{\rm t}(\zeta)$.

\section{The method}

\subsection{A more general view on the problem}

We first relax the assumption of isotropic turbulence used in the Sect.~\ref{Intro}
(but will later return to it).
We remain however with the definition of mean fields by averaging over all $x$ and $y$.
Then, as already roughly indicated above, $\meanB_x$ and $\meanB_y$ may only depend on $z$ and time
but $\meanB_z$, because of $\nab \cdot \meanBB = 0$, must be independent of $z$.
Furthermore all first--order spatial derivatives of $\meanBB$
can be expressed by the components of $\nab \times \meanBB$, that is, of $\meanJJ$,
where $\meanJ_z = 0$.
Instead of \eq{meanEMFzzprime2} we have then
\EQ
\meanemf_i (z) = \int \left[ \hat{\alpha}_{ij} (\zeta) \meanB_j (z - \zeta)
   - \hat{\eta}_{ij} (\zeta) \meanJ_j (z - \zeta) \right] \, \dd \zeta
\label{eq11}
\EN
and instead of \eq{meanEMFzzprime3}
\EQ
\tilde{\meanemf}_i (k) = \tilde{\alpha}_{ij} (k) \tilde{\meanB}_j (k)
   - \tilde{\eta}_{ij} (k) \tilde{\meanJ}_j (k) \, ,
\label{eq13}
\EN
with real $\hat{\alpha}_{ij} (\zeta)$ and $\hat{\eta}_{ij} (\zeta)$, which are even in $\zeta$,
and real $\tilde{\alpha}_{ij} (k)$ and $\tilde{\eta}_{ij} (k)$, which are even in $k$.
A justification of these relations is given in \App{justific}.
We have further
\EQ
\tilde{\alpha}_{ij} (k) = \int \hat{\alpha}_{ij} (\zeta) \cos k \zeta \, \dd \zeta \, , \quad
    \tilde{\eta}_{ij} (k) = \int \hat{\eta}_{ij} (\zeta) \cos k \zeta \, \dd \zeta \, .
\label{eq15}
\EN
Since $\meanJ_3 = 0$ the $\hat{\eta}_{i3}$, as well as the $\tilde{\eta}_{i3}$,
are of no interest.

In the following we restrict attention to $\meanemf_x$ and $\meanemf_y$
and assume that $\meanB_z$ is equal to zero.
We note that $\meanemf_z$ and the contributions of $\meanB_z$ to $\meanemf_x$ and $\meanemf_y$
are anyway without interest for the mean--field induction equation,
which contains $\meanEMF$ only in the form $\nab \times \meanEMF$,
that is, they do not affect the evolution of $\meanBB$.
We may formulate the above restriction in a slightly different way by saying
that we consider in the following $\meanemf_i$, $\alpha_{ij}$ and $\eta_{ij}$
as well as $\tilde{\meanemf}_i$, $\tilde{\alpha}_{ij}$ and $\tilde{\eta}_{ij}$
only for $1 \leq i, j \leq 2$.

As for the mean-field treatment of a Roberts flow depending on $x$ and $y$
only (and not on $z$) we refer to the aforementioned studies
(R\"adler et al.\ 2002a,b, R\"adler \& Brandenburg 2003).
Following the ideas explained there we may conclude
that $\hat{\alpha}_{ij} (\zeta)= \hat{\alpha} (\zeta) \delta_{ij}$
and $\hat{\eta}_{ij} (\zeta) = \hat{\eta}_{\rm t} (\zeta) \delta_{ij}$
with functions $\hat{\alpha}$ and $\hat{\eta}_{\rm t}$ of $\zeta$,
and analogously $\tilde{\alpha}_{ij} (k)= \tilde{\alpha} (k) \delta_{ij}$
and $\tilde{\eta}_{ij} (k) = \tilde{\eta}_{\rm t} (k) \delta_{ij}$
with functions $\tilde{\alpha}$ and $\tilde{\eta}_{\rm t}$ of $k$,
all for $1 \leq i, j \leq 2$.
For obvious reasons the same is true for homogeneous isotropic turbulence.

\subsection{Test field method}

We calculate the $\tilde{\alpha}_{ij} (k)$ and $\tilde{\eta}_{ij} (k)$,
or $\tilde{\alpha} (k)$ and $\tilde{\eta} (k)$, numerically by employing
the test field method of Schrinner et al.\ (2005, 2007).
It has been originally developed for the calculation
of the full $\alpha$ and $\eta$ tensors [in the sense of \eq{meanemfi}]
for convection in a spherical shell.
Brandenburg (2005) employed this method to obtain results for stratified shear flow turbulence
in a local cartesian domain using the shearing sheet approximation.
More recently, Sur et al.\ (2008) calculated in this way the dependences of $\alpha$ and $\eta_{\rm t}$
for isotropic turbulence on the magnetic Reynolds number,
and Brandenburg et al.\ (2008) have calculated the magnetic diffusivity
tensor for rotating and shear flow turbulence.
However, in all these cases no nonlocality in the connection between $\meanEMF$ and $\meanBB$
has been taken into account.

Following the idea of Schrinner et al.\ we first derive expressions for $\meanEMF$
with several specific $\meanBB$, which we call ``test fields".
We denote the latter by $\meanBB^{p q}$ and define%
\footnote{The notation used here differs slightly from that in Brandenburg et al.\ (2008),
where first test fields $\meanBB^p$ were introduced and only later the two versions
$\meanBB^{p \rm{c}}$ and $\meanBB^{p \rm{s}}$ are considered.}
\begin{eqnarray}
\meanBB^{1 \, \rm{c}} &=& B \, (\cos k z , 0, 0) \, , \quad
    \meanBB^{2 \, \rm{c}} = B \, (0, \cos k z, 0) \, ,
\nonumber\\
\meanBB^{1 \, \rm{s}} &=& B \, (\sin k z , 0, 0) \, , \quad
    \meanBB^{2 \, \rm{s}} = B \, (0, \sin k z, 0) \, ,
\label{eq21}
\end{eqnarray}
with any constant $B$ and any fixed value of $k$.
We then replace $\meanBB$ and $\meanJJ$ in \eq{eq11}
by $\meanBB^{p \, \rm{c}}$ and $\nab \times \meanBB^{p \, \rm{c}}$
or by $\meanBB^{p \, \rm{s}}$ and $\nab \times \meanBB^{p \, \rm{s}}$.
Denoting the corresponding $\meanEMF$ by $\meanEMF^{p \, \rm{c}}$
or by $\meanEMF^{p \, \rm{s}}$, respectively,
and using \eq{eq15} we find
\begin{eqnarray}
\meanemf_i^{p \, \rm{c}} (z)  &=&  B \, \left[ \tilde{\alpha}_{ip} (k) \cos k z
    - \tilde{\eta}^\dag_{ip} (k) \, k \sin k z \right] \, ,
\nonumber\\
\meanemf_i^{p \, \rm{s}} (z)  &=&  B \, \left[ \tilde{\alpha}_{ip} (k) \sin k z
    + \tilde{\eta}^\dag_{ip} (k) \, k \cos k z \right] \, ,
\label{eq23}
\end{eqnarray}
for $1 \leq i, p \leq 2$, where
\EQ
\tilde{\eta}^\dag_{ip} = \tilde{\eta}_{il} \epsilon_{lp3}
    = \pmatrix{- \tilde{\eta}_{12} & \tilde{\eta}_{11} \cr - \tilde{\eta}_{22} & \tilde{\eta}_{21}} \, .
\label{eq25}
\EN
From this we conclude
\begin{eqnarray}
\tilde{\alpha}_{ij} (k)  &=&  B^{-1} \left[ \meanemf_i^{j \, \rm{c}} (z) \cos k z
    + \meanemf_i^{j \, \rm{s}} (z) \sin k z \right] \, ,
\nonumber\\
\tilde{\eta}^\dag_{ij} (k) &=& - (k B)^{-1} \left[ \meanemf_i^{j \, \rm{c}} (z) \sin k z
    - \meanemf_i^{j \, \rm{s}} (z) \cos k z \right]
\label{eq27}
\end{eqnarray}
for $1 \leq i, j \leq 2$.

These relations allow us to calculate the $\tilde{\alpha}_{ij}$ and $\tilde{\eta}_{ij}$
if the $\meanemf_i^{pq}$ with $1 \leq i, p \leq 2$ for both  $q = {\rm c}$ and $q = {\rm s}$ are known.
In preparing the numerical calculation we start from the induction equation.
Its uncurled form reads
\EQ
{\partial\AAA\over\partial t}=\UU\times\BB-\eta\JJ,
\label{dAAA}
\EN
where $\AAA$ is the magnetic vector potential, $\BB=\nab\times\AAA$,
and $\JJ=\nab\times\BB$.
Here the Weyl gauge of $\AAA$ is used.
Taking the average of \eq{dAAA} we obtain
\EQ
{\partial\meanAA\over\partial t}=\meanUU\times\meanBB+\overline{\uu\times\bb}-\eta\meanJJ \, .
\label{dAAAm}
\EN
From \eq{dAAA} and \eq{dAAAm} we conclude
\EQ
{\partial\aaa\over\partial t}=\meanUU\times\bb+\uu\times\meanBB
+\uu\times\bb-\overline{\uu\times\bb}-\eta\jj \, ,
\label{eq31}
\EN
where $\aaa = \AAA - \meanAA$ and $\jj=\JJ -\meanJJ=\nab\times\bb$.

For the calculation of the $\meanEMF^{pq}$ we are interested in the $\bb^{pq}=\nab\times\aaa^{pq}$
which occur in response to the test fields $\meanBB^{pq}$.
Specifying \eq{eq31} in that sense we obtain\footnote{In the corresponding expression
(27) of Brandenburg (2005) the $\meanUU$ term is incorrect.
This did not affect his results because $\meanUU=0$.}
\EQ
{\partial\aaa^{pq}\over\partial t}=\meanUU\times\bb^{pq}+\uu\times\meanBB^{pq}
+\uu\times\bb^{pq}-\overline{\uu\times\bb^{pq}}
-\eta\jj^{pq}.
\label{nonSOCA}
\EN
Equations of this type are called ``test field equations".

So far no approximation such as the second order
correlation approximation (SOCA), also known as first order smoothing
approximation, has been made.
If we were to make this assumption, terms that are nonlinear in the fluctuations
would be neglected and \eq{nonSOCA} would simplify to
\EQ
{\partial\aaa^{pq}\over\partial t}=\meanUU\times\bb^{pq}+\uu\times\meanBB^{pq}
-\eta\jj^{pq}\quad\mbox{(for SOCA only)}.
\label{SOCA}
\EN
In the following SOCA results will be shown in some cases for comparison only.

In either of the two cases the $\tilde{\alpha}_{ij}$ and $\tilde{\eta}_{ij}$
are to be calculated from $\meanEMF^{pq}=\overline{\uu\times\bb^{pq}}$.
More details of the numerical calculations of the $\meanEMF^{pq}$
will be given below in \Sec{simul}.

Returning once more to \eq{eq27} we note that the $\meanEMF^{pq}$ depend
on both $k$ and $z$ introduced with the $\meanBB^{pq}$.
As a consequence of imperfect simulations of the turbulence they may also depend on the time $t$.
The $\tilde{\alpha}_{ij}$ and $\tilde{\eta}_{ij}$ however should depend on $k$
but no longer on $z$ and $t$.
We remove the latter dependences of our results by averaging
$\tilde{\alpha}_{ij}$ and $\tilde{\eta}_{ij}$ over $z$ and $t$.
For the Roberts flow there should be no such $z$ or $t$ dependences.

The relations \eq{eq27} allow the determination of all components
of $\tilde{\alpha}_{ij}$ and $\tilde{\eta}_{ij}$ with $1 \leq i,j \leq 2$.
We know already that $\tilde{\alpha}_{ij} = \tilde{\alpha} \delta_{ij}$
and $\tilde{\eta}_{ij} = \tilde{\eta}_{\rm t} \delta_{ij}$,
that is, $\tilde{\alpha}_{11} = \tilde{\alpha}_{22} = \tilde{\alpha}$,
$\tilde{\eta}_{11} = \tilde{\eta}_{22} = \tilde{\eta}$
and $\tilde{\alpha}_{12} = \tilde{\alpha}_{21} = \tilde{\eta}_{12} = \tilde{\eta}_{21} = 0$.
We may therefore determine $\tilde{\alpha}$ and $\tilde{\eta}_{\rm t}$
according to $\tilde{\alpha} = \tilde{\alpha}_{11}$ and $\tilde{\eta}_{\rm t} = \tilde{\eta}_{11}$
by using the two test fields $\meanBB^{1 \, q}$ and the relations \eq{eq27} with $i = j = 1$ only.

\subsection{Flow fields}

\subsubsection{Roberts flow}

We consider here a special form of a steady flow
which, in view of its dynamo action, has already been studied by Roberts (1972).
It has no mean part, $\meanUU = {\bf 0}$, and $\uu$ is given by
\EQ
\uu = - \zz \times \nab \psi + k_{\rm f} \psi \zz \, ,
\label{UUrobflow}
\EN
where
\EQ
\psi = (u_0/k_0) \cos k_0x \, \cos k_0y \, , \quad k_{\rm f}=\sqrt{2}\,k_0
\label{UUrobflow2}
\EN
with some constant $k_0$.
The flow is fully helical, $\nab\times\uu=k_{\rm f}\uu$.
The component form of $\uu$ as defined by \eq{UUrobflow} and  \eq{UUrobflow2} reads
\EQ
\uu=u_0 \big(-\cos k_0x\,\sin k_0y, \sin k_0x\,\cos k_0y, \sqrt{2}\cos k_0x\,\cos k_0y \big).
\label{RobertsFlow}
\EN
We note that $\overline{\uu^2} = u_0^2$.

\subsubsection{Turbulence}

Next, we consider isotropic, weakly compressible turbulence and use an
isothermal equation of state with constant speed of sound, $c_{\rm s}$.
Considering first the full velocity field $\UU = \meanUU + \uu$
we thus accept the momentum equation in the form
\EQ
{\partial\UU\over\partial t}=-\UU\cdot\nab\UU-c_{\rm s}^2\nab\ln\rho
+\ff+\rho^{-1}\nab\cdot2\rho\nu\SSSS,
\label{dUU}
\EN
where $\ff$ is a random forcing function consisting of circularly
polarized plane waves with positive helicity and random direction, and
${\sf S}_{ij}={1\over2}(U_{i,j}+U_{j,i})-{1\over3}\delta_{ij}\nab\cdot\UU$
is the traceless rate of strain tensor.
The forcing function is chosen such that the magnitude of the wavevectors,
$|\kk_{\rm f}|$, is in a narrow interval around an average value,
which is simply denoted by $k_{\rm f}$.
The corresponding scale, $k_{\rm f}^{-1}$, is also referred to as the
outer scale or the energy-carrying scale of the turbulence.
More details concerning the forcing function are given in the appendix of
Brandenburg \& Subramanian (2005b).
With the intention to study the mean electromotive force in the purely kinematic limit
the Lorentz force has been ignored.

In addition to the momentum equation we use the continuity equation in the form
\EQ
{\partial\ln\rho\over\partial t}=-\UU\cdot\nab\ln\rho-\nab\cdot\UU.
\label{dlnrho}
\EN

In all simulations presented in this paper the strength of the forcing is adjusted such
that the flow remains clearly subsonic,
that is, the mean Mach number remains below 0.2.
Hence for all practical purposes the flow can be considered incompressible
(Dobler et al.\ 2003).
In these simulations
no mean flow develops, that is $\meanUU={\bf 0}$, so $\UU=\uu$.

\subsection{Simulations}
\label{simul}

The relevant equations are solved in a computational domain of size $L \times L \times L$
using periodic boundary conditions.
In the case of the Roberts flow \eq{RobertsFlow} we fix $L$ by  $L=2\pi/k_0$.
Only four of the test field equations \eq{nonSOCA}
(those with $p=1$, $q = {\rm c}$ and $q = {\rm s}$)
are solved numerically.
With turbulence in the kinematic regime the four equations \eq{dUU} and \eq{dlnrho}
for $\UU$ and $\ln\rho$ are solved together with four
of the test field equations \eq{nonSOCA}
(again with $p=1$, $q = {\rm c}$ and $q = {\rm s}$).

Due to the finiteness of the domain in $z$ direction and the periodic boundary conditions,
quantities like $\meanEMF$ and $\meanBB$ have to be considered as functions that are periodic in $z$.
The Fourier integrals used for representing these quantities,
$Q(z) = \int \tilde{Q}(k) \exp(\ii k z) \, \dd (k/2 \pi)$,
turn into Fourier series,
$Q(z) = \sum Q_n \exp(\ii k_n z)/L$,
where $k_n=2\pi n/L$ and the summation is over $n = 0, \pm 1, \pm 2, \ldots$.
By this reason
only discrete values of $k$, that is $k = k_n$, are admissible in \eq{eq13}--\eq{eq27}.
In this framework we may determine the $\tilde{\alpha}$ and $\tilde{\eta}_{\rm t}$
only for these $k_n$.

As explained above, the test field procedure yields
$\tilde{\alpha}$ and $\tilde{\eta}_{\rm t}$
not as functions of $k$ alone.
They may also show some dependence on $z$ and $t$.
After having averaged over $z$, time averages are then taken over
a suitable stretch of the full time series where these averages are
approximately steady.
We use the time series further to calculate error bars as the maximum
departure between these averages and the averages obtained from one of
three equally long subsections of the full time series.

In all cases the simulations have been carried out using the
{\sc Pencil Code}\footnote{\url{http://www.nordita.org/software/pencil-code}}
which is a high-order finite-difference code (sixth order in space and third
order in time) for solving the compressible hydromagnetic equations
together with the test field equations.
In the case of the Roberts flow, of course, only the test field equations
are being solved.

\section{Results}

\subsection{Roberts flow}

Let us first recall some findings of earlier work,
which are presented, e.g., by R\"adler (2002a,b).
We use here the definitions
\EQ
\alpha_0 = - \half u_0 \, , \quad \eta_{\rm t0} = \half u_0 / k_{\rm f} \, , \quad
     \Rm = u_0 / \eta k_{\rm f} \, .
\label{eq52}
\EN
Adapting the results of analytic calculations in the framework of SOCA
to the assumptions and notations of this paper (see \App{robdyn}) we have
\EQ
\alpha / \alpha_0 = \eta_{\rm t} / \eta_{\rm t0} = \Rm \, .
\label{eq51}
\EN
Moreover, in the general case, also beyond SOCA, it was found that
\EQ
\alpha = \alpha_0 \Rm \, \phi (\sqrt{2} \Rm)
\label{eq53}
\EN
with a function $\phi$ satisfying $\phi (0) = 1$ and vanishing with growing argument.
This function has been calculated numerically and is plotted, e.g., in R\"adler et al.\ (2002a,b).

\begin{figure}[t!]
\centering\includegraphics[width=\columnwidth]{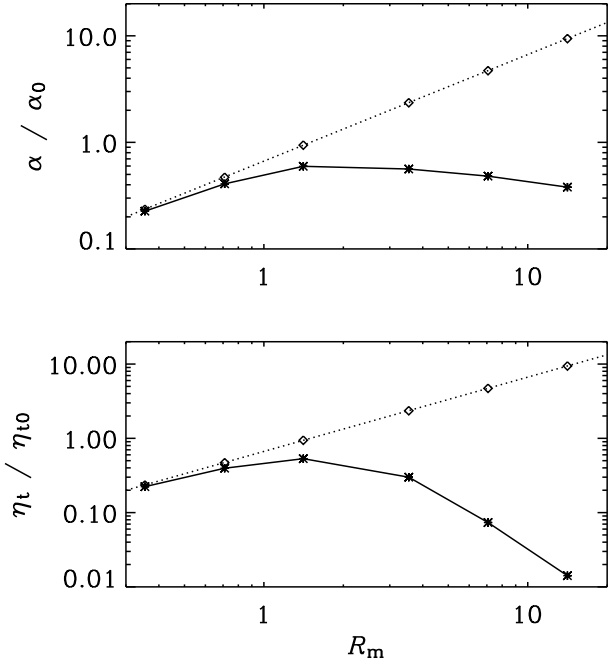}\caption{
Dependences of the normalized
$\alpha$ and $\eta_{\rm t}$ on $\Rm$ for the Roberts flow
in the general case, i.e.\ independent of SOCA (solid lines), and in SOCA (dotted lines).
}\label{palp_robflow}\end{figure}

\Fig{palp_robflow} shows results for $\alpha$ and $\eta_{\rm t}$ obtained
both by general test field calculations using \eq{nonSOCA} and under the restriction to SOCA using \eq{SOCA}.
These results for $\alpha$ agree completely with both \eq{eq51} and \eq{eq53},
and those for $\eta_{\rm t}$ agree completely with \eq{eq51}.
Unfortunately we have no analytical results for $\eta_{\rm t}$ beyond SOCA.

Proceeding now to the $\tilde{\alpha} (k)$ and $\tilde{\eta}_{\rm t} (k)$ we first note that in SOCA,
as shown in \App{calc},
\EQ
\tilde\alpha(k)= \frac{\alpha_0 \Rm}{1+(k/k_{\rm f})^2} \, , \quad
    \tilde\eta_{\rm t}(k)= \frac{\eta_{\rm t0}\Rm}{1+(k/k_{\rm f})^2} \, .
\label{eq55}
\EN
The corresponding $\hat{\alpha} (\zeta)$ and $\hat{\eta}_{\rm t} (\zeta)$, again in SOCA, read
\EQ
\hat{\alpha} (\zeta) = \half \alpha_0 k_{\rm f} \Rm \exp(- k_{\rm f} |\zeta|) \, , \quad
    \hat{\eta}_{\rm t} (\zeta) = \half \eta_{\rm t0} k_{\rm f} \Rm \exp(- k_{\rm f} |\zeta|) \, .
\label{eq57}
\EN

\begin{figure}[t!]
\centering\includegraphics[width=\columnwidth]{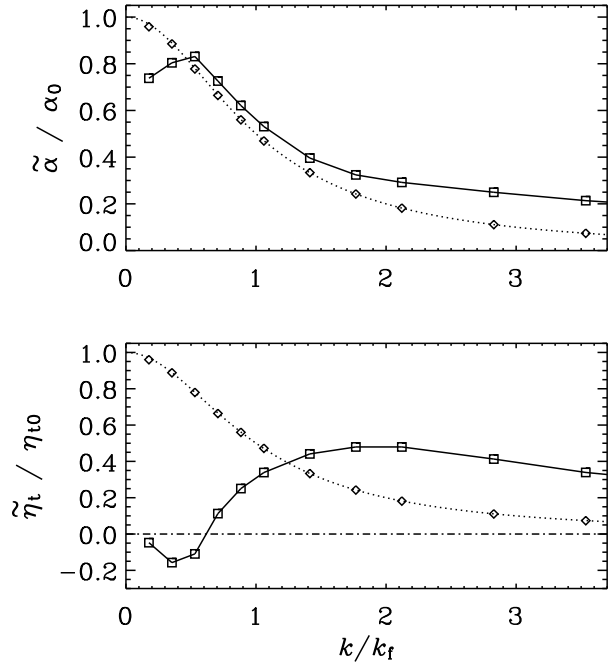}\caption{
Dependences of the normalized
$\tilde{\alpha}$ and $\tilde{\eta}_{\rm t}$ on $k/k_{\rm f}$
for the Roberts flow with $\Rm=10/\sqrt{2}\approx7.1$ (solid lines),
compared with normalized SOCA results for $\tilde{\alpha} / \Rm$ and $\tilde{\eta}_{\rm t} / \Rm$,
which are independent of $\Rm$ (dotted lines).
}\label{ppk_rob}\end{figure}

In \Fig{ppk_rob} results of test field calculations for the functions
$\tilde{\alpha} (k)$ and $\tilde{\eta}_{\rm t} (k)$ with $\Rm=10 / \sqrt{2} \approx 7.1$ are shown.
We note that $\tilde{\eta}_{\rm t}$ becomes negative for small $k$.
The same has been observed with another but similar flow of Roberts type
(R\"adler \& Brandenburg 2003).
For comparison, SOCA
results obtained in two different ways are also shown:
those according to the analytic relations \eq{eq55}
and those calculated numerically by the test field method with \eq{SOCA}.
Both agree very well with each other.

\begin{figure}[t!]
\centering\includegraphics[width=\columnwidth]{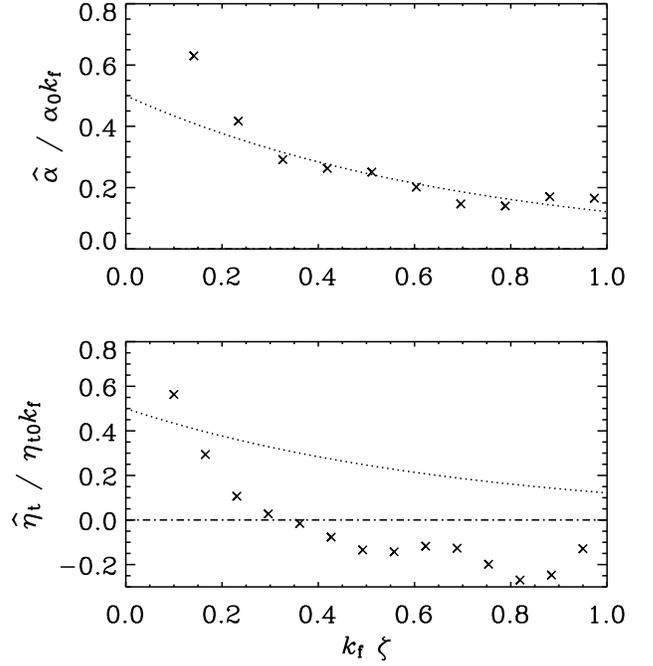}\caption{
Normalized integral kernels $\hat\alpha$ and $\hat\eta_{\rm t}$ versus $k_{\rm f} \zeta$
for the Roberts flow with $\Rm=10/\sqrt{2}\approx7.1$ (solid lines),
compared with normalized SOCA results for $\hat{\alpha} / \Rm$ and $\hat{\eta}_{\rm t} / \Rm$,
which are independent of $\Rm$ (dotted lines).
The full width half maximum values of $k_{\rm f} \zeta$ for $\hat{\alpha}$ and $\hat\eta_{\rm t}$
are about $0.5$ and $0.2$, respectively.
}\label{ppk_rob_kernel}\end{figure}

In order to obtain the results for the kernels $\hat\alpha (\zeta)$ and
$\hat\eta_{\rm t} (\zeta)$ we have
calculated integrals as in \eq{eq15} numerically
using the data plotted in \Fig{ppk_rob}.
The results are represented in \Fig{ppk_rob_kernel}.
Again, analytical and numerical SOCA results are shown for comparison.
Note that the profiles
of $\hat\alpha (\zeta)$ and $\hat\eta_{\rm t} (\zeta)$ beyond SOCA
are rather narrow compared with those under SOCA,
and that of $\hat\eta_{\rm t} (\zeta)$ even more narrow than that of $\hat\alpha (\zeta)$.

\subsection{Isotropic turbulence}

Results for homogeneous isotropic turbulence have been obtained by solving
the hydrodynamic equations \eqs{dUU}{dlnrho}
simultaneously with the test field equations \eq{nonSOCA}
in a domain of size $L^3$.
The forcing wavenumbers $k_{\rm f}$ are fixed by  $k_{\rm f}/k_1=5$ and $10$.
Instead of the definitions \eq{eq52} we use now
\EQ
\alpha_0 = - \onethird u_{\rm rms} \, , \quad \eta_{\rm t0} = \onethird u_{\rm rms} / k_{\rm f} \, , \quad
     \Rm = u_{\rm rms} / \eta k_{\rm f} \, .
\label{eq59}
\EN

Within this framework the dependence of $\alpha$ and $\eta_{\rm t}$ on $\Rm$
has been studied by Sur et al.\ (2008).
They considered two cases, one with $\nu/\eta=0.1$ and another one with $u_{\rm rms}/\nu k_{\rm f}=2.2$.
Remarkably, they found that $\alpha/\alpha_0$ and $\eta/\eta_0$ approach unity
for $\Rm \gg 1$.

\begin{figure}[t!]
\centering\includegraphics[width=\columnwidth]{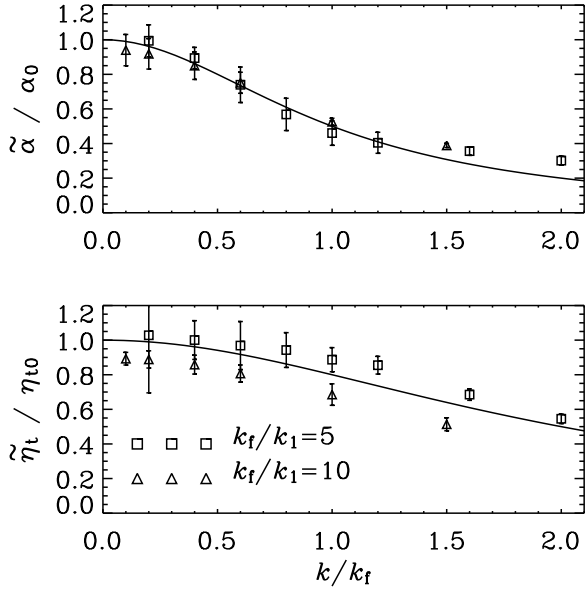}\caption{
Dependences of the normalized $\tilde\alpha$ and $\tilde\eta_{\rm t}$
on the normalized wavenumber $k/k_{\rm f}$
for isotropic turbulence forced at wavenumbers $k_{\rm f}/k_1=5$ with $\Rm=10$ (squares)
and $k_{\rm f}/k_1=10$ with $\Rm=3.5$ (triangles), all with $\nu/\eta=1$.
The solid lines give the Lorentzian fits \eq{KernelsTurb}.
}\label{ppk}\end{figure}

\FFig{ppk} shows results for $\tilde\alpha (k)$ and $\tilde\eta_{\rm t} (k)$
with $\nu / \eta = 1$.
Both $\tilde\alpha$ and $\tilde\eta_{\rm t}$ decrease monotonously with increasing $|k|$.
The two values of $\tilde\alpha$ for a given $k/k_{\rm f}$ but different $k_{\rm f}/k_1$ and $\Rm$
are always very close together.
The functions $\tilde\alpha (k)$ and $\tilde\eta_{\rm t} (k)$
are well represented by Lorentzian fits of the form
\EQ
\tilde\alpha(k)={\alpha_0\over1+(k/k_{\rm f})^2} \, ,\quad
\tilde\eta_{\rm t}(k)={\eta_{\rm t0}\over1+(k/2k_{\rm f})^2} \, .
\label{KernelsTurb}
\EN

\begin{figure}[t!]
\centering\includegraphics[width=\columnwidth]{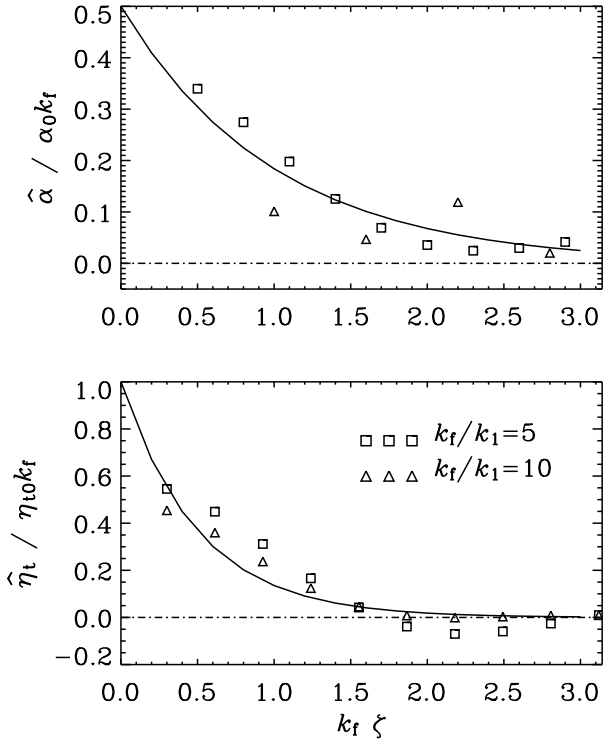}\caption{
Normalized integral kernels $\hat\alpha$ and $\hat\eta_{\rm t}$ versus
$k_{\rm f}\zeta$ for isotropic turbulence forced at wavenumbers $k_{\rm f}/k_1=5$ with $\Rm=10$ (squares)
and $k_{\rm f}/k_1=10$ with $\Rm=3.5$ (triangles), all with $\nu/\eta=1$.
The solid lines are defined by \eq{KernelsTurb2}.
}\label{ppk_kernel}\end{figure}

In \Fig{ppk_kernel} the kernels $\hat\alpha (\zeta)$ and $\hat\eta_{\rm t} (\zeta)$,
again with $\nu / \eta = 1$,
obtained by calculating numerically integrals as in \eq{eq15}, are depicted.
Also shown are the Fourier transforms of the Lorentzian fits,
\EQ
\hat\alpha(\zeta)=\half\alpha_0 k_{\rm f} \exp(-k_{\rm f}|\zeta|) \, ,\quad
\hat\eta_{\rm t}(\zeta)=\eta_{\rm t0} k_{\rm f} \exp(-2k_{\rm f}|\zeta|) \, .
\label{KernelsTurb2}
\EN
Evidently, the profile of $\hat\eta_{\rm t}$ is half as wide as that of $\hat\alpha$.
This corresponds qualitatively to our observation with the Roberts flow beyond SOCA,
see the crosses in \Fig{ppk_rob_kernel}.
There is however no counterpart to the negative values of $\hat\eta_{\rm t}$
that occur in the example of the Roberts flow.

The results presented in \Figs{ppk}{ppk_kernel} show no noticeable dependences on $\Rm$.
Although we have not performed any systematic survey in $\Rm$,
we interpret this as an extension of the above--mentioned results of Sur et al.\ (2008)
for $\alpha$ and $\eta_{\rm t}$ to the integral kernels $\hat\alpha$
and $\hat\eta_{\rm t}$.
Of course, this should to be checked also with larger values of $\Rm$.
Particularly interesting would be a confirmation of different widths
of the profiles of $\hat\alpha$ and $\hat\eta_{\rm t}$.

\section{Discussion}

Our results are important for calculating mean--field dynamo models.
The mean--field induction equation governing $\meanBB$, here defined as average over $x$ and $y$,
with $\meanEMF$ according to \eq{meanEMFzzprime2}, allows solutions of the form
$\Re \left[ \meanBB_0 \exp (\ii k z + \lambda t) \right]$, $\meanB_{0z} = 0$, with
the growth rate
\begin{equation}
\lambda = - \left[\eta + \tilde{\eta}_{\rm t} (k) \right] k^2 \pm \tilde{\alpha} (k) k \, .
\label{eq71}
\end{equation}
A dynamo occurs if $\lambda$ is non--negative.
Since $\tilde{\alpha} \leq 0$ in all examples considered this occurs with the lower sign,
and we focus attention on this case only.
In the limit of a local connection between $\meanEMF$ and $\meanBB$ the
$\tilde{\eta}_{\rm t} (k)$ and $\tilde{\alpha} (k)$
turn into $\tilde{\eta}_{\rm t} (0)$ and $\tilde{\alpha} (0)$, respectively.

When using the definitions \eq{eq52} for the Roberts flow, or \eq{eq59} for isotropic turbulence,
we may write \eq{eq71} in the form
\EQ
\lambda = \eta_{\rm t0} k_{\rm f}^2 \left\{
 -\left[{\gamma\over\Rm} + {\tilde{\eta}_{\rm t} (k/k_{\rm f} )\over \eta_{\rm t0}}\right] {k\over k_{\rm f}}
+{\tilde{\alpha} (k/k_{\rm f}) \over \alpha_0} \right\} \, \frac{k}{k_{\rm f}} \, ,
\label{eq73}
\EN
where $\gamma = 2$ for the Roberts case and $\gamma = 3$ for the isotropic case.
Since $\tilde{\eta}_{\rm t}$ and $\tilde{\alpha}$ depend only via $k/k_{\rm f}$ on $k$
we have chosen the arguments $k/k_{\rm f}$.

Consider first the Roberts flow, that is \eq{eq73} with $\gamma = 2$.
Clearly $\lambda$ is non--negative in some interval $0 \leq k/k_{\rm f} \leq \kappa_0$
and it takes there a maximum.
Dynamos with $k/k_{\rm f} > \kappa_0$ are impossible.
Of course, $\kappa_0$ depends on $\Rm$.
With the analytic SOCA results \eq{eq55} we find
that $\kappa_0 = \half \Rm^2$ for small $\Rm$ and that it grows monotonically with $\Rm$
and approaches unity in the limit of large $\Rm$.
For small $\Rm$ (to which the applicability of SOCA is restricted) a dynamo can work only
with small $k/k_{\rm f}$, that is, with scales of the mean magnetic field that are much smaller than
the size of a flow cell.
Furthermore, $\kappa_0$ never exceeds the corresponding values for vanishing nonlocal effect,
which is $\half \Rm^2 / (1 + \half \Rm^2)$.
In that sense the nonlocal effect favors smaller $k$, that is, larger scales of the mean magnetic field.
With the numerical results beyond SOCA represented in \Fig{ppk_rob},
with $\Rm=10 / \sqrt{2}$, we have $\kappa_0 \approx 0.90 ... 0.95$, again a value smaller than unity.
In this case, too, a dynamo does not work with scales of the mean magnetic field smaller than that of a flow cell.
There is no crucial impact of the negative values of $\tilde{\eta}_{\rm t}$
for $k/k_{\rm f} < 0.8$ on the dynamo.

\begin{figure}[t!]
\centering\includegraphics[width=\columnwidth]{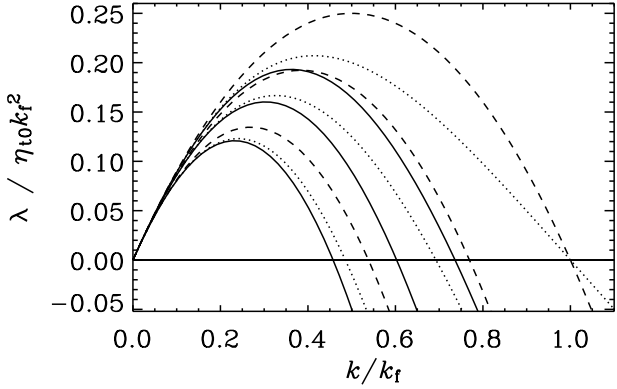}\caption{
Normalized growth rate $\lambda(k)$
for isotropic turbulence, calculated according to relation \eq{eq73} with $\gamma = 3$
and with $\tilde{\eta}_{\rm t}/\eta_{\rm t0}$ and $\tilde{\alpha}/\alpha_0$ as given in \eq{KernelsTurb},
for $\Rm \to \infty$ (upper solid line), as well as $\Rm = 10$ and 3.5
(next lower solid lines).
For comparison $\lambda$ is also shown for the case in which $\tilde{\eta}_{\rm t}/\eta_{\rm t0}$
coincides with $\tilde{\alpha}/\alpha_0$ as given in \eq{KernelsTurb} (dotted lines)
and for that of vanishing nonlocal effect,
in which $\tilde{\eta}_{\rm t}/\eta_{\rm t0} = \tilde{\alpha}/\alpha_0 = 1$ (dashed lines),
each for the same three values of $\Rm$.
}\label{plamk1}\end{figure}

Proceed now to isotropic turbulence and consider \eq{eq73} with $\gamma = 3$.
Again $\lambda$ is non--zero in the interval $0 \leq k/k_{\rm f} \leq \kappa_0$
and it takes there a maximum.
Some more details are shown in \Fig{plamk1}.
With the Lorentzian fits \eq{KernelsTurb} of the results depicted in \Fig{ppk}
we find $\kappa_0 \approx 0.60$ for $\Rm=10$ (and 0.45 for $\Rm=3.5$).
In the limit of vanishing nonlocal effects
it turns out that $\kappa_0 \approx 0.82$ for $\Rm=10$ (and 0.59 for $\Rm=3.5$).
We have to conclude that dynamos are only possible if the scale of the mean magnetic field clearly exceeds
the outer scale of the turbulence.
In addition we see again that the nonlocal effect favors smaller $k$,
or larger scales of the mean magnetic field.

These findings may become an important issue especially for nonlinear
dynamos or for dynamos with boundaries.
Examples of the last kind were studied,
e.g., by Brandenburg \& Sokoloff (2002)
and Brandenburg \& K\"apyl\"a (2007).
In these cases however
the underlying turbulence is no longer homogeneous and therefore
the kernels $\hat\alpha$ and $\hat\eta_{\rm t}$
are no longer invariant under translations, that is,
depend not only on $\zeta$ but also on $z$.
The finite widths of the $\hat\alpha$ and $\hat\eta_{\rm t}$ kernels may
be particularly important if there is also shear, because then there
can be a travelling dynamo wave that may also show strong gradients
in the nonlinear regime (Stix 1972, Brandenburg et al.\ 2001).

\begin{figure}[t!]
\centering\includegraphics[width=\columnwidth]{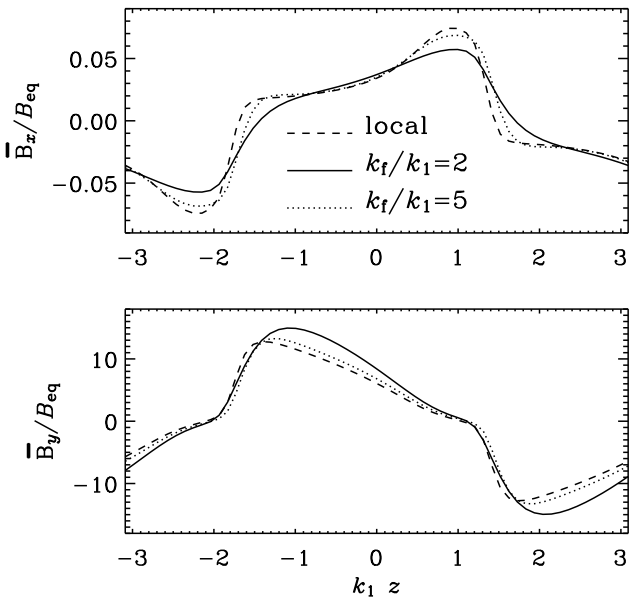}\caption{
Mean magnetic field components $\meanB_x$ and $\meanB_y$,
normalized by the equipartition field strength $B_{\rm eq}$, in the one-dimensional nonlinear
dynamo model characterized in the text, for different values of $k_{\rm f}/ k_1$
and for vanishing nonlocal effects.
}\label{pcomp_nloc}\end{figure}

For another illustration of the significance of a finite width
of the kernels $\hat\alpha$ and $\hat\eta_{\rm t}$
we consider a one-dimensional nonlinear mean--field model
with periodic boundary conditions.
We modify here the model of Brandenburg et al.\ (2001, Sect.~6),
with a dynamo number of 10 (corresponding to 5 times supercritical)
and $\Rm=25$,
by introducing the integral kernels \eq{KernelsTurb2}.
\FFig{pcomp_nloc} shows the components of the mean magnetic field
for two different values of $k_{\rm f}/k_1$ and for the conventional case
where the kernels are delta--functions.
Note that $k_1$ corresponds to the largest scale of the magnetic field
compatible with the boundary condition.
It turns out that the magnetic field profiles are not drastically altered
by the nonlocal effect.
Small values of $k_{\rm f}/k_1$, however, correspond to smoother profiles.

Let us start again from $\meanEMF$ in the form \eq{meanEMFzzprime2}, specify there,
in view of isotropic turbulence, $\hat{\alpha}$ and $\hat{\eta}_{\rm t}$ according to \eq{KernelsTurb2},
and represent $\meanBB (z - \zeta)$ and $\meanJJ (z - \zeta)$ by Taylor series with respect to $\zeta$.
A straightforward evaluation of the integrals provides us then with
\EQ
\meanEMF (z) = \sum_{n \geq 0}
    \Big( \frac{\alpha_0}{k_{\rm f}^{2n}} \frac{\partial^{2n} \meanBB (z)}{\partial z^{2n}}
    - \frac{\eta_{\rm t0}}{(2 k_{\rm f})^{2n}} \frac{\partial^{2n} \meanJJ (z)}{\partial z^{2n}} \Big) \, .
\label{hyperdiff}
\EN
This corresponds to relations of the type \eq{meanemfi} or \eq{meanEMF},
simply generalized by taking into account higher than first--order derivatives of $\meanBB$.

The terms with derivatives of $\meanJJ$ in \eq{hyperdiff} can be interpreted in the sense of hyperdiffusion.
While all of them have the same signs in real space, the signs of the corresponding terms
in Fourier space alternate, which implies that every second term acts in an anti-diffusive manner.
Thus, a truncation of the expansion should only be done such that the last remaining term has an even $n$,
as otherwise anti-diffusion would dominate on small length scales
and cause $\meanBB$ to grow  beyond any bound.

There are several investigations in various fields in which hyperdiffusion has been considered.
In the purely hydrodynamic context, R\"udiger (1982) derived
a hyperviscosity term and showed that this improves the
representation of the mean velocity profile in turbulent channel flows.
In the context of passive scalar diffusion, Miesch et al.\ (2000) determined
the hyperdiffusion coefficients for turbulent convection and found that
they scale with $n$ like in \Eq{hyperdiff}.
We are however not aware of earlier studies differentiating between diffusive
and anti-diffusive terms.

We have investigated the nonlocal cases presented in \Fig{pcomp_nloc}
using truncations of the expansion \eq{hyperdiff}.
However, two problems emerged.
Firstly, terms with higher derivatives produce Gibbs phenomena,
i.e.\ wiggles in $\meanBB$, so the results in \Fig{pcomp_nloc} are not
well reproduced.
Secondly, high--order hyperdiffusion terms tend to give severe constraints
on the maximum admissible time step, making this approach
computationally less attractive.
It appears therefore that a direct evaluation of the convolution terms
is most effective.

\section{Conclusions}

The test field procedure turned out
to be a robust method for determining turbulent transport coefficients
(see Brandenburg 2005, Sur et al.\ 2008 and Brandenburg et al.\ 2008).
The present paper shows that this also applies to the
Fourier transforms of the integral kernels
which occur in the nonlocal connection between mean electromotive force and mean magnetic field,
in other words to the more general scale--dependent version of those transport coefficients.
For isotropic turbulence the kernels $\hat\alpha$ and $\hat\eta_{\rm t}$ have a
dominant large-scale part and decline monotonously with increasing wavenumbers.
This is consistent with earlier findings (cf.\ Brandenburg \& Sokoloff 2002),
where however the functional form of the decline remained rather uncertain.
Our present results suggest exponential kernels, corresponding to Lorentzian profiles
in wavenumber space.
The kernel for the turbulent magnetic diffusivity is about half as wide
as that for the alpha effect.
This result is somewhat unexpected and would be worthwhile to confirm
before applying it to more refined mean field models.
On the other hand, the effects of nonlocality become really important
only when the scale of the magnetic field variations is comparable
or smaller than the outer scale of the turbulence.

One of the areas where future research of nonlocal turbulent transport
coefficients is warranted is thermal convection.
Here the vertical length scale of the turbulent plumes is often
comparable to the vertical extent of the domain.
Earlier studies by Miesch et al.\ (2000) on turbulent thermal convection
confirmed that the transport of passive scalars is nonlocal, but it is
also more advective than diffusive.
It may therefore be important to also allow for nonlocality in time.
This would make the expansion of passive scalar perturbations more
wave-like, as was show by Brandenburg et al.\ (2004) using forced
turbulence simulations.

\begin{acknowledgements}
We acknowledge the allocation of computing resources provided by the
Centers for Scientific Computing in Denmark (DCSC), Finland (CSC),
and Sweden (PDC).
We thank Matthias Rheinhardt for stimulating discussions.
A part of the work reported here was done during stays of K.-H.\ R.\
and M.\ S.\ at NORDITA.
They are grateful for NORDITA's hospitality.
\end{acknowledgements}

\appendix

\section{Justification of equations \eq{meanEMFzzprime} and \eq{eq11}}
\label{justific}

In view of \eq{meanEMFzzprime}
we start with equation \eq{kernel} for $\meanEMF$, put $\meanEMF_0 = {\bf 0}$,
assume that $\KK \circ \meanBB$ is a purely spatial convolution.
Applying then the Fourier transform as defined
by $Q (\xx) = \int \tilde{Q} (\kk) \exp (\ii \kk \cdot \xx)\,\dd^3 (k / 2 \pi)$
we obtain
\EQ
\tilde{\meanemf}_i (\kk) = \tilde{K}_{ij} (\kk) \tilde{\meanB}_j (\kk) \, .
\label{J01}
\EN
Since $\meanEMF$ and $\meanBB$ have to be real we conclude that
$\tilde{K}^*_{ij} (\kk) = \tilde{K}_{ij} (-\kk)$.
Further the assumption of isotropic turbulence requires that $\tilde{K}_{ij}$
is an isotropic tensor.
We write therefore
\EQ
\tilde{K}_{ij} = \tilde{\alpha} (k) \delta_{ij} + \tilde{\alpha}' (k) k_i k_j
    + \ii \tilde{\eta}_{\rm t}(k) \epsilon_{ijk} k_k
\label{J03}
\EN
with $\tilde{\alpha}$, $\tilde{\alpha}'$ and $\tilde{\eta}_{\rm t}$ being real functions of $k = |\kk|$.
Considering further that $\kk \cdot \tilde{\meanBB} = 0$ and $\ii \kk \times \tilde{\meanBB} = \tilde{\meanJJ}$
we find
\EQ
\tilde{\meanEMF}(\kk) = \tilde{\alpha}(k) \tilde{\meanBB}(\kk) - \tilde{\eta}_{\rm t} (k) \tilde{\meanJJ} (\kk)  \, .
\label{J05}
\EN
Transforming this in the physical space we obtain immediately \eq{meanEMFzzprime}.

In view of \eq{eq11}
we start again from equation \eq{kernel} and put $\meanEMF_0 = {\bf 0}$
but we have to consider $\KK \circ \meanBB$ now as a convolution with respect to $z$ only.
Applying a Fourier transformation defined by
$Q (z) = \int \tilde{Q} (k) \exp (\ii k z)\,\dd (k / 2 \pi)$
we obtain a relation analogous to \eq{J01},
\EQ
\tilde{\meanemf}_i (k) = \tilde{K}_{ij} (k) \tilde{\meanB}_j (k) \, .
\label{J07}
\EN
and may now conclude that $\tilde{K}^*_{ij} (k) = \tilde{K}_{ij} (-k)$.
We arrive so at
\EQ
\tilde{K}_{ij} = \tilde{\alpha}_{ij} (k) + \ii k \tilde{\eta}'_{ij} (k)
\label{J09}
\EN
with real tensors $\tilde{\alpha}_{ij}$ and $\tilde{\eta}'_{ij}$, which are even in $k$.
Combining \eq{J07} and \eq{J09} and considering that the $\ii k \tilde{\meanB}_i$
can be expressed by the $\tilde{\meanJ}_i$
($\ii k \tilde{\meanB}_1 = \tilde{\meanJ}_2$, $\ii k \tilde{\meanB}_2 = - \tilde{\meanJ}_1$,
$\ii k \tilde{\meanB}_3 = 0$)
we may confirm first \eq{eq13} and so also \eq{eq11}.

\section{Mean--field results for the Roberts flow}
\label{robdyn}

A mean--field theory of the Roberts dynamo,
developed in view of the Karlsruhe dynamo experiment,
has been presented, e.g., in papers by R\"adler et al.\ (2002a,b),
in the following referred to as R02a and R02b.
There a fluid flow like that given by (\ref{RobertsFlow}) is considered
but without coupling of its magnitudes in the $xy$--plane and in the $z$--direction.
The mean fields are defined by averaging over finite areas in the $xy$--plane
so that they may still depend on $x$ and $y$ in addition to $z$.
As shown in the mentioned papers $\meanEMF$, when contributions
with higher than first--order derivatives of $\meanBB$ are ignored,
has then the form
\begin{eqnarray}
\meanEMF &=& - \alpha_\perp \big[ \meanBB - (\zz \cdot \meanBB) \zz \big]
    - \beta_\perp \nab \times \meanBB - (\beta_\parallel
    - \beta_\perp) \big[ \zz \cdot (\nab \times \meanBB) \big] \zz
\nonumber\\
&& - \beta_3 \zz \times \big[ \nab (\zz \meanBB) + (\zz \cdot \nab) \meanBB \big]
\label{A01}
\end{eqnarray}
with constant coefficients $\alpha_\perp$, $\beta_\perp$, $\beta_\parallel$ and $\beta_3$
[see (R02a 9) or (R02b 9)].
Reducing this to the case considered above, in which $\meanBB$ depends no longer
on $x$ and $y$, we find
\EQ
\meanEMF = \alpha \big[ \meanBB - (\zz \cdot \meanBB) \zz \big]
    - \eta_{\rm t} \nab \times \meanBB \, ,
\label{A03}
\EN
where $\nab\times\meanBB = \zz\times\partial\meanBB/\partial z$, and
\EQ
\alpha = - \alpha_\perp \, , \quad \eta_{\rm t} = \beta_\perp + \beta_3 \, .
\label{A05}
\EN

Results for $\alpha_\perp$, $\beta_\perp$, $\beta_\parallel$ and $\beta_3$
obtained in the second--order correlation approximation are given in (R02a 19)
and (R02a 49) as well as in (R02b 19) and (R02b 38).
When fitting them with $u_\perp = (2/\pi) u_0$, $u_\parallel = \sqrt{2} (2/\pi)^2 u_0$,
$\pi/a = k_1$, $\sqrt{2} \pi/a = k_{\rm f}$,
${\Rm}_\perp = \sqrt{2} \Rm$ and ${\Rm}_\parallel = (8/\pi) \Rm$
to the assumptions and notations used above we find just \eq{eq51}.
Likewise (R02a 20) and also (R02b 20) lead to \eq{eq53}.

\section{$\tilde{\alpha}$ and $\tilde{\eta}_{\rm t}$ under SOCA for Roberts flow}
\label{calc}

Let us start with the relation \eq{A03} and subject it to a Fourier transformation
so that
\EQ
\tilde{\meanEMF} = \overline{\uu \times \tilde{\bb}}
    = \tilde{\alpha} \, \big[ \tilde{\meanBB} - (\zz \cdot \tilde{\meanBB}) \zz \big]
    - \ii k \, \tilde{\eta}_{\rm t} \, \zz \times \tilde{\meanBB} \, .
\label{B01}
\EN
From the induction equation we have
\EQ
\eta (\nab^2 - k^2) \tilde{\bb} = - \, (\nab + \ii k \zz) \times (\uu \times \tilde{\meanBB})
\, , \quad k \tilde{b}_z = 0
 \, .
\label{B03}
\EN
The solution $\tilde{\bb}$ reads
\begin{eqnarray}
\tilde{\bb} &=& - \frac{1}{k^2 + k_{\rm f}^2} \Big\{ \zz \times \nab (\tilde{\meanBB} \cdot \nab \psi)
   - k_{\rm f} (\tilde{\meanBB} \cdot \nab \psi) \zz
\nonumber\\
&& \qquad \qquad + \ii k \big[ \zz \times \nab \psi \, (\zz \cdot \tilde{\meanBB})
   + k_{\rm f} \psi (\tilde{\meanBB} - (\zz \cdot \tilde{\meanBB}) \zz) \big] \Big\} \, ,
\label{B05}
\end{eqnarray}
where $\uu$ is according to \eq{UUrobflow} expressed by $\psi$.
Calculate now $\tilde{\meanemf}_x$ or $\tilde{\meanemf}_y$
and note that $\overline{\psi^2} = \textstyle{1\over4} (u_0 / k_1)^2$
and $\overline{(\partial \psi / \partial x)^2} = - \, \overline{\psi \partial^2 \psi / \partial x^2}
= \textstyle{1\over4} u_0^2$.
When comparing the result with \eq{B01} we find immediately \eq{eq55}.
Using then relations of the type \eq{eq15} we find also \eq{eq57}.

\end{document}

%% file: aa.tex
\newcommand{\EQ}{\begin{equation}}
\newcommand{\EN}{\end{equation}}
\newcommand{\EQA}{\begin{eqnarray}}
\newcommand{\ENA}{\end{eqnarray}}
\newcommand{\eq}[1]{(\ref{#1})}

\newcommand{\Eq}[1]{Eq.~(\ref{#1})}

\newcommand{\eqs}[2]{(\ref{#1}) and~(\ref{#2})}

\newcommand{\App}[1]{Appendix~\ref{#1}}
\newcommand{\Sec}[1]{Sect.~\ref{#1}}

\newcommand{\Fig}[1]{Fig.~\ref{#1}}
\newcommand{\FFig}[1]{Figure~\ref{#1}}

\newcommand{\Figs}[2]{Figs~\ref{#1} and \ref{#2}}

\newcommand{\meanEMF}{\overline{\vec{\cal E}}}
\newcommand{\meanemf}{\overline{\cal E}}

\newcommand{\meanB}{\overline{B}}

\newcommand{\meanJ}{\overline{J}}

\newcommand{\meanAA}{\overline{\vec{A}}}
\newcommand{\meanBB}{\overline{\vec{B}}}
\newcommand{\meanJJ}{\overline{\vec{J}}}
\newcommand{\meanUU}{\overline{\vec{U}}}

{}
{}
{}
{}
{}
{}

\newcommand{\zz}{\hat{\mbox{\boldmath $z$}} {}}
\newcommand{\xx}{{\vec{x}}}

\newcommand{\UU}{{\vec{U}}}
\newcommand{\uu}{{\vec{u}}}
\newcommand{\BB}{{\vec{B}}}
\newcommand{\JJ}{{\vec{J}}}
\newcommand{\jj}{{\vec{j}}}
\newcommand{\AAA}{{\vec{A}}}
\newcommand{\aaa}{{\vec{a}}}
\newcommand{\bb}{{\vec{b}}}

\newcommand{\ff}{\mbox{\boldmath $f$} {}}

\newcommand{\KK}{\mbox{\boldmath $K$} {}}

\newcommand{\kk}{{\vec{k}}}

\newcommand{\nab}{\vec{\nabla}}

\newcommand{\xxi}{\vec{\xi}}

\newcommand{\SSSS}{\mbox{\boldmath ${\sf S}$} {}}

\newcommand{\ii}{{\rm i}}

\newcommand{\dd}{{\rm d} {}}

\def\Rm{R_{\rm m}}

\def\half{{\textstyle{1\over2}}}

\def\onethird{{\textstyle{1\over3}}}

\newcommand{\yapj}[3]{ #1, {ApJ,} {#2}, #3}

\newcommand{\yan}[3]{ #1, {AN,} {#2}, #3}

\newcommand{\yana}[3]{ #1, {A\&A,} {#2}, #3}

\newcommand{\ygafd}[3]{ #1, {GApFD,} {#2}, #3}

\newcommand{\ypf}[3]{ #1, {PhFl,} {#2}, #3}

\newcommand{\yprl}[3]{ #1, {PRL,} {#2}, #3}
\newcommand{\ypre}[3]{ #1, {PRE,} {#2}, #3}

\newcommand{\yptrs}[3]{ #1, {Phil. Trans. Roy. Soc.,} {#2}, #3}
\newcommand{\ymn}[3]{ #1, {MNRAS,} {#2}, #3}

\newcommand{\yjour}[4]{ #1, {#2}, {#3}, #4}

\newcommand{\ybook}[3]{ #1, {#2} (#3)}
\newcommand{\yproc}[5]{ #1, in {#3}, ed. #4 (#5), #2}

\newcommand{\pmn}[1]{ #1, {MNRAS} (in press)}